\documentclass[manuscript]{aastex631}
\usepackage{CJK}
\usepackage{graphicx}
\usepackage{subfigure}
\usepackage{booktabs}
\usepackage{multirow}
\usepackage{amssymb}

\begin{document}
\begin{CJK*}{UTF8}{gbsn}

\title{Multiwavelength Properties of Infrared-Faint Radio Sources Based on Spectral Energy Distribution Analysis}

\correspondingauthor{Yihang Zhang, Lulu Fan}
\email{zhang$\_$yh@mail.ustc.edu.cn, llfan@ustc.edu.cn}

\author[0009-0003-9423-2397]{Yihang Zhang (张迤航)}
\affiliation{Department of Astronomy,
University of Science and Technology of China, Hefei 230026, China}
\affiliation{School of Astronomy and Space Science, University of Science and Technology of China, Hefei 230026, China}

\author[0000-0003-4200-4432]{Lulu Fan (范璐璐)}
\affiliation{Department of Astronomy,
University of Science and Technology of China, Hefei 230026, China}
\affiliation{School of Astronomy and Space Science, University of Science and Technology of China, Hefei 230026, China}
\affiliation{Deep Space Exploration Laboratory, Hefei 230088, China}

\author[0000-0003-4341-0029]{Tao An (安涛)} 
\affiliation{Key Laboratory of Radio Astronomy, Shanghai Astronomical Observatory, 80 Nandan Road, 200030, Shanghai, China}

\author[0000-0002-2322-5232]{Jun Yang (杨军)}
\affiliation{Department of Space, Earth and Environment, Chalmers University of Technology, Onsala Space Observatory, SE-439 92 Onsala, Sweden}

\author[0009-0004-7885-5882]{Weibin Sun (孙卫斌)}
\affiliation{Department of Astronomy,
University of Science and Technology of China, Hefei 230026, China}
\affiliation{School of Astronomy and Space Science, University of Science and Technology of China, Hefei 230026, China}

\author[0009-0008-1319-498X]{Haoran Yu (于浩然)}
\affiliation{Department of Astronomy,
University of Science and Technology of China, Hefei 230026, China}
\affiliation{School of Astronomy and Space Science, University of Science and Technology of China, Hefei 230026, China}

\author[0000-0002-2547-0434]{Yunkun Han (韩云坤)} 
\affiliation{Yunnan Observatories, Chinese Academy of Sciences, 396 Yangfangwang, Guandu District, Kunming, 650216, P. R. China}
\affiliation{Center for Astronomical Mega-Science, Chinese Academy of Sciences, 20A Datun Road, Chaoyang District, Beijing, 100012, P. R. China}
\affiliation{Key Laboratory for the Structure and Evolution of Celestial Objects, Chinese Academy of Sciences, 396 Yangfangwang, Guandu District, Kunming, 650216, P. R. China}
\affiliation{International Centre of Supernovae, Yunnan Key Laboratory, Kunming 650216, P. R. China}

\begin{abstract}
Infrared-faint radio sources (IFRSs) are believed to be a rare class of radio-loud active galactic nuclei (RL AGN) characterized by their high radio-to-infrared flux density ratios of up to several thousands. Previous studies have shown that a fraction of IFRSs are likely to be hosted in dust-obscured galaxies (DOGs). In this paper, our aim was to probe the dust properties, star formation rate (SFR), and AGN activity of IFRSs by modeling the UV-to-infrared spectral energy distribution (SED) of 20 IFRSs with spectroscopic redshifts ranging from 1.2 to 3.7. We compare the Bayesian evidence of a three-component model (stellar, AGN and cold dust) with that of a two-component model (stellar and cold dust) for six IFRSs in our sample with far-infrared (FIR) photometry and find that the three-component model has significantly higher Bayesian evidence, suggesting that IFRSs are most likely to be AGN. The median SED of our IFRS sample shows similarities to AGN-starburst composite in the IR regime. The derived IR luminosities of IFRSs indicate that they are low-luminosity counterparts of high-redshift radio galaxies. We disentangle the contributions of AGN-heated and star-formation-heated dust to the IR luminosity of IFRSs and find that our sample is likely AGN-dominated. However, despite the evidence for significant impact of AGN on the host galaxy, the AGN luminosity of our sample does not show correlation with the SFR of the sources. 

\end{abstract}


\section{Introduction} \label{sec:intro}

Infrared-faint radio sources (IFRSs) are rare galaxies that are bright at radio wavelengths but extremely faint in the near-infrared regime. They were first discovered by \cite{Norris2006} as radio sources detected at $\lambda \ =\ 20$ cm in the deep radio observations of the Australia Telescope Large Area Survey (ATLAS) in Chandra Deep Field South (CDFS) but without counterparts in the Spitzer Wide-area Infrared Extragalactic Survey (SWIRE) at $\lambda \ =\ 3.6, 4.5, 5.8, 8.0$ and $24\ \mu m$. This discovery was quite surprising, since it is generally assumed that active galactic nuclei (AGN) and star-forming galaxies (SFGs) detectable at $\lambda \ =\ 20$ cm would also be detectable by SWIRE. 31 more sources were found in the European Large Area IR Space Observatory Survey South 1 (ELAIS-S1) field\citep{middelberg2008}. The IFRSs found in these deep fields have 1.4 GHz flux density range from a few hundred $\mu$Jy to a few tens of mJy, while non-detection in the SWIRE imposes an upper flux limit of 5 $\mu$Jy in the 3.6 $\mu$m band. In particular, all 53 IFRS found in CDFS and ELAIS-S1 lack optical counterparts. 

Early research characterized IFRSs as sources without detectable counterpart at any \textit{Spitzer} wavelength, which is a rather loose definition and depends on the sensitivity of the survey \citep{Norris2006}. \citet{Zinn2011} proposed a new set of IFRS selection criteria independent of surveys. A source is identified as IFRS if the two following conditions are met: ({\romannumeral 1}) Flux density ratio $S_{\rm 20cm}/S_{\rm 3.6\mu m}\ >\ 500$; ({\romannumeral 2}) $S_{\rm 3.6\mu m}\ <30 \mu Jy$. The first criterion selects sources with extreme radio-to-IR flux density ratios, excluding contamination from SFGs. The second criterion ensures that low-redshift AGNs are not selected, as they typically exhibit a relatively high flux density in the IR. \cite{Zinn2011}'s criteria enable the search of a larger, brighter IFRS population with optical and infrared detections. Using these criteria, \cite{Collier2014} searched for IFRSs using the Unified Radio Catalog \citep[URC;][]{kimball2008unified} and the \textit{Wide-field Infrared Survey Explorer} \citep[\textit{WISE};][]{wright2010wise} data and compiled a sample of 1317 IFRSs, which is the largest IFRS sample by far. 
 
Previous investigations of IFRSs that focused on radio wavelengths suggested that they are young RL AGNs. Very Long Baseline Interferometry (VLBI) observations of some IFRSs showed brightness temperatures about $10^6$ K, indicating non-thermal emission from AGN \citep{Norris2007VLBI,middelberg2008b,herzog2015active}. Most VLBI-detected IFRSs show compact cores, suggesting that they contain young AGNs whose jets have not expanded yet \citep{herzog2015active}. A substantial fraction of IFRSs are compact steep-spectrum (CSS) and GHz-Peaked-Spectrum (GPS) sources \citep{middelberg2011radio,Collier2014, Herzog2016}. GPS sources exhibit a spectrum turnover at $\sim$ 1 GHz and are considered to represent the earliest evolutionary stage of AGN \citep{randall2011unbiased}. CSS sources are compact radio sources with a spectral peak at $\sim$ 100 MHz and a steep spectral index across the GHz range ($\alpha \ \leq \ -0.8$). CSS sources are more extended than GPS sources and possibly represent an intermediate evolutionary phase between GPS sources and the larger Fanaroff–Riley type \uppercase\expandafter{\romannumeral 1}/\uppercase\expandafter{\romannumeral 2} (FR \uppercase\expandafter{\romannumeral 1}/ FR \uppercase\expandafter{\romannumeral 2}) galaxies. The IFRS identified in the ELAIS-S1 field have steep radio spectra with a median index of $\alpha$ = -1.4 between 2.3 GHz and 8.4 GHz \citep{middelberg2011radio}, which is steeper than the general radio source population ($\alpha$ = -0.86) and the AGN source population ($\alpha$ = -0.82) in ELAIS-S1, further supporting that a substantial fraction of IFRS contain young AGN. 

Using ultra-deep \textit{Spitzer} imaging, IR counterparts of some IFRSs were detected, while upper flux limits in the IRAC bands for nondetected sources were attained through image stacking. IR-to-radio SED modeling of IFRSs showed that 3C sources can reproduce the data when redshifted to z $>$ 2 \citep{GA2008, huynh2010}. \cite{herzog2015dust} performed \textit{Herschel} maps stacking to obtain FIR flux density upper limits of 6 IFRSs. Combined with radio detections and SWIRE flux density upper limits, their SED modeling found that known RL quasar templates failed to match the photometric constraints of IFRSs, but RL quasar or CSS sources at lower redshifts (z $\leq$ 5) with additional dust obscuration could reproduce the IR-faintness of IFRSs. Spectroscopic redshifts of IFRSs are crucial for accurate SED modeling and decomposition. Due to their faintness in the optical and IR regimes, it has been difficult to measure the spectroscopic redshifts of IFRSs. The first spectroscopic redshifts of IFRSs were presented by \cite{Collier2014} and \cite{herzog2014infrared}, who identified the redshifts of 21 IFRSs within the range $1.8 \leq z \leq 3.0$. These sources with measured spectroscopic redshifts lie at the IR-bright end of IFRSs, with IR flux densities between 14 $\mu$Jy and 30 $\mu$Jy at 3.6 $\mu$m or 3.4 $\mu$m, implying that the IR-fainter ones are located at even higher redshifts. \cite{Orenstein2019} presented the largest sample of IFRSs with spectroscopic redshifts, containing 131 sources with a median redshift of $z\ =\ 2.68$. 

The flux density ratio $S_{\rm 1.4GHz}/S_{\rm 3.6\mu m}$ of IFRSs substantially overlaps with that of high redshift radio galaxies (HzRGs), ranging from several hundred to several thousand, which is uncommon for general radio sources \citep{Norris2011,seymour2007massive}. HzRGs are a class of powerful radio galaxies ($L_{\rm 3GHz}\ >\ 10^{26}$ W Hz$^{-1}$) at high redshifts (1 $\leq$ z $\leq$ 5). They are believed to be the progenitors of the most massive galaxies in the local universe \citep{seymour2007massive,debreuk2010hzrg}. A potential link between IFRSs and HzRGs has been suggested, as HzRGs are the only high redshift population known to exhibit the same extreme radio-to-IR flux density ratios. \cite{herzog2015dust} found that the SEDs of their IFRS sample could be explained by HzRG templates at even higher redshifts (z $>$ 5) or with additional dust extinction. Besides these similarities, a notable difference between IFRSs and HzRGs is that HzRGs are known to host highly accreting AGN and vigorous star formation, both of which contribute to their high mid- and far-IR luminosities. IFRSs could be fainter, higher-redshift siblings of HzRGs. SED modeling of IFRS constrained by spectroscopic redshift and optical-to-IR photometric data could uncover the nature of IFRS and test the possible relation between IFRS and HzRG. 

In this paper, we present a redshift-based optical-to-IR SED modeling of 20 IFRSs. Using a Bayesian approach, we decompose the emissions from different components and find strong evidence for AGN in IFRSs. In Section \ref{sec:data}, we describe our sample selection and the construction of an optical-to-IR SED. In Section \ref{sec:method}, we introduce our Bayesian method for SED modeling. We present our results of SED modeling in Section \ref{sec:results}. In Section \ref{sec:discussion}, we discuss the parameters derived from the SED modeling and the possible nature of IFRSs. We summarize our conclusions in Section \ref{sec:summary}.

\section{Sample Selection and Multiwavelength Data} \label{sec:data}

\subsection{Sample selection}
\citet{Collier2014} slightly modified \citet{Zinn2011}'s criteria by replacing the 3.6 $\mu m$ flux density with the 3.4 $\mu m$ flux density from \textit{WISE} \citep{wright2010wise}, which enables the construction of a significantly larger IFRS sample.

To build a sample for SED modeling, we started with the 167 IFRSs that had been identified with spectroscopic redshifts\citep{Collier2014,Herzog2016,Singh2017,Orenstein2019}. We obtained the 20 cm radio data for 167 IFRSs from the Unified Radio Catalog (URC) version 2.0 compiled by \citet{kimball2013updated}. This radio catalog contains data from the NRAO VLA Sky Survey \citep[NVSS;][]{condon1998nvss}, Faint Images of the Radio Sky at Twenty Centimeters \citep[FIRST;][]{becker1995first}, Green Bank 6cm survey \citep[GB6;][]{gregory1996gb6}, the Westerbork Northern Sky Survey \citep[WENSS;][]{rengelink1997westerbork}, and the Sloan Digital Sky Survey Data Release 9 \citep[SDSS DR9][]{ahn2012sdssdr9}. The 3.4 $\mu$m and 3.6 $\mu$m data are from the ALLWISE data release \citep{cutri2021allwise} and \textit{Spitzer} Enhanced Imaging Products \citep[SEIP;][]{SEIPsource}, respectively. We applied the following criteria to select the sample discussed in this paper.\\
({\romannumeral 1})We cross-matched the NVSS positions to ALLWISE and SEIP with a search radius of 5 arcsec and applied a signal-to-noise ratio limit of SNR $\ge 5$ in the W1 band and the IRAC band 1\\
({\romannumeral 2})We selected all sources with $S_{\rm 20cm}/S_{\rm 3.4-3.6\mu m}\ \ge \ 500$\\
({\romannumeral 3})We selected all sources with $S_{\rm 3.6\mu m}\ \le 30\ \mu Jy$\\

\begin{figure}[!htbp]
    \centering
    \includegraphics[width=1.0\textwidth]{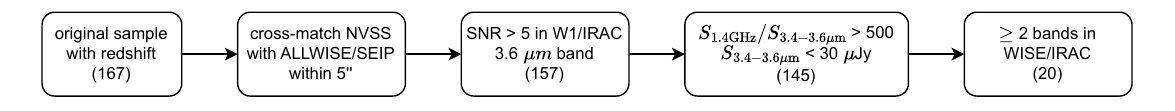}
    \caption{Flowchart of our sample selection process. The number in parentheses indicate the number of sources remaining after each step of selection.}
    \label{fig:flowchart}
\end{figure}

After applying the selection criteria mentioned above, our final sample consisted of 145 IFRSs with spectroscopic redshifts. To construct the UV-to-IR SEDs for IFRSs, we retrieved multiwavelength photometry from various catalogs, which will be discussed in the next section. 

\subsection{Infrared data}
The W1, W2, and W3 photometry for the IFRS sample is from the ALLWISE data release \citep{cutri2021allwise}. To convert the catalog Vega magnitudes to flux density in Janskys (Jy), we used a flat color correction factor determined for IFRSs by \citet{Collier2014}, which are
\begin{equation}
  \ S_{\rm W1}=306.682\times10^{-M_{\rm W1}/2.5}\  Jy
\end{equation}
\begin{equation}
  \ S_{\rm W2}=170.663\times10^{-M_{\rm W2}/2.5}\  Jy
\end{equation}
\begin{equation}
  \ S_{\rm W3}=29.045\times10^{-M_{\rm W3}/2.5}\  Jy
\end{equation}
32 IFRSs in our sample have near-infrared (NIR) broad-band photometric data from different surveys, including the eighth data release of the United Kingdom Infrared Telescope (UKIRT) Infrared Deep Sky Survey \citep[UKIDSS DR8;][]{Lawrence2012UKIDSS} in J, H, and K bands and the Visible and Infrared Survey Telescope for Astronomy (VISTA) Deep Extragalactic Observations Survey \citep[VIDEO;][]{Jarvis2012VISTA}{}{} in Z, Y, J, H, and $K_s$ bands. 15 IFRSs in our sample have mid-infrared (MIR) data from SEIP \citep[SEIP;][]{SEIPsource} Source List, which includes data from the four bands of the Infrared Array Camera \citep[IRAC;][]{fazio2004irac} and the 24 $\mu$ m band of the Multiband Imaging Photometer \citep[MIPS;][]{rieke2004mips}. For far-infrared (FIR) photometry, 6 IFRSs in our sample have \textit{Herschel} Photoconductor Array Camera and Spectrometer \citep[PACS;][]{poglitsch2010PACS} observations at 70 and 160 $\mu$m and Spectral and Photometric Imaging REceiver \citep[SPIRE;][]{griffin2010SPIRE} observations at 250, 350 and 500 $\mu$m, respectively. The IR photometry of the six sources is listed in Table \ref{tab:photo}.

\subsection{Optical data}
129 IFRSs in our sample have optical counterparts in SDSS Data Release 17 \citep{accetta2022sdss17}, Canada-France-Hawaii Telescope Legacy Survey \citep[CFHTLS;][]{hudelot2012cfht}, and the Multi-wavelength Survey by Yale-Chile optical imaging in ECDF-S \citep[MUSYC;][]{cardamone2010musyc}.

\begin{deluxetable*}{lcccccccccc}
\tablecaption{IR photometry of 6 IFRSs}
\label{tab:photo}
\tabletypesize{\scriptsize}
\setlength{\tabcolsep}{2pt}
\tablehead
{\\
\colhead{Source ID} & 
\colhead{$3.6\ \mu m^a$} & 
\colhead{$4.5\ \mu m^a$} & 
\colhead{$5.8\ \mu m^a$} & 
\colhead{$8\ \mu m^a$} & 
\colhead{$24\ \mu m^a$ } & 
\colhead{$100\ \mu m^b$}&
\colhead{$160\ \mu m^b$}&
\colhead{$250~ \mu m^b$ } &
\colhead{$350~ \mu m^b$ } &
\colhead{$500~ \mu m^b$ } \\ 
  &
\colhead{$\mu Jy$} & 
\colhead{$\mu Jy$} & 
\colhead{$\mu Jy$} & 
\colhead{$\mu Jy$} & 
\colhead{mJy}&
\colhead{mJy}&
\colhead{mJy} &
\colhead{mJy} &
\colhead{mJy} &
\colhead{mJy} \\}
\startdata
122 & 42.33$\pm$0.37 &61.88$\pm$0.58 &92.28$\pm$2.01 & 166.8$\pm$3.56 & 0.57$\pm$0.04 &...&...& 33.2$\pm$7.60 & 44.0$\pm$9.3 & 30.1$\pm$8.6\\
136 & 11.04$\pm$0.21 & 16.15$\pm$0.71 & 35.85$\pm$8.08 & 181.40$\pm$10.55 & 13.4$\pm$7.7&13.4$\pm$7.71&38.98$\pm$10.72 &5.6$\pm$2.3 & 12.1$\pm$2.5 & 9.0$\pm$3.4\\
154 & 4.76$\pm$0.08 & 4.928$\pm$0.09 & 5.485$\pm$0.32 & 2.89$\pm$0.3 & 0.18$\pm$0.01&...&... & 11.7$\pm$3.8 & 13.2$\pm$3.8 & 10.6$\pm$4.0\\
160 & 27.34$\pm$0.11 & 50.46$\pm$0.19 & 110.3$\pm$0.64 & 243.9$\pm$1.23 & 0.87$\pm$0.01&...&... & 16.1$\pm$2.4 & 14.6$\pm$2.2 & 6.7$\pm$2.6\\
161 & 11.4$\pm$0.55 & 11.96$\pm$0.88 & 48.59$\pm$5.04 &...& 0.59$\pm$0.05 &...&... & 3.8$\pm$2.4 & 7.5$\pm$4.5 &...\\
162 & 13.4$\pm$0.99 & 13.47$\pm$0.14 & 18.56$\pm$0.52 & 21.86$\pm$0.9 & 0.35$\pm$0.01&...&... & 14.8$\pm$1.1 & 15.1$\pm$1.0 & 10.9$\pm$1.4\\
\enddata
\tablecomments{$^a$ 3.6, 4.5, 5.8, 8.0 and 24 $\mu m$ photometry from \textit{Spitzer} SEIP catalog \citep{SEIPsource}.\\ $^b$100, 160, 250, 350, and 500 $\mu m$ photometry from \textit{Herschel} Multi-tiered Extragalactic Survey \citep{hermes2012}. }
\end{deluxetable*}

In order to obtain reliable estimations on the properties of IFRS, we selected 20 IFRSs with counterparts in at least two bands in ALLWISE or SEIP Source List from the original sample for SED modeling. Their redshift, 20 cm, and 3.4/3.6 $\mu$m data are listed in Table \ref{tab:subsample}. To assess the representativeness of our selected sample of 20 IFRSs, we conducted a Kolmogorov-Smirnov (K-S) test on both the redshift distributions and the radio-to-IR flux density ratios. The K-S test for redshift distributions and radio-to-IR flux density ratios resulted in p-values of 0.066 and 0.105, respectively, which are marginally above the commonly used significance threshold of 0.05. This suggests a potential, though not statistically significant, difference in the redshift and radio-to-IR flux density ratio distributions between the selected and original samples. The K-S tests imply that there may be a slight bias in our selection, which will be considered when interpreting the results of the SED analysis for the selected IFRS sample.

\begin{deluxetable}{lccccc}
\tablecaption{The sample of 20 IFRSs with spectroscopic redshift}
\label{tab:subsample}
\tabletypesize{\scriptsize}
\setlength{\tabcolsep}{2pt}
\tablehead
{\\
\colhead{Source ID} & 
\colhead{RA} & 
\colhead{Dec} & 
\colhead{redshift$^a$}&
\colhead{$S_{\rm 20cm}^b$}&
\colhead{$S_{\rm 3.4/3.6\mu m}^c$}\\
 &
\colhead{(J2000)} & 
\colhead{(J2000)} & 
 &
\colhead{(mJy)}&
\colhead{($\mu Jy$)}\\}
\startdata
6&152.63407 &8.13474 &2.33 &20.49 &27.28\\
37&205.87463 	&32.13341 	&3.15 &19.54 &23.31\\
41&241.50610 	&29.53084 	&2.18 &16.88 &28.02\\
61&245.94765 	&54.71742 	&2.23 &22.76 &25.98\\
74&178.61794 	&14.16786 	&2.68 &14.70 &29.26\\ 
83&191.42380 	&25.98835 	&2.26 &15.02 &29.78\\
106&249.94554 	&40.65931 	&2.63 &19.75 &21.67\\
110&143.87793 	&61.28998 	&1.42 &45.30 &28.28\\
116&202.34253 	&5.33735 	&2.99 &44.90 &29.89\\
119&208.90853 	&6.18864 	&2.39 &31.10 &28.36\\
122&219.59085 	&34.66693 	&2.34 &12.26 &20.36\\
133&249.57249 	&41.45825 	&2.23 &36.81 &26.93\\
136&34.51425 	&-5.64040 	&3.57 &8.91 &14.28\\
154&53.31985 	&-28.00457 &2.64 &4.63 &7.40\\
160&34.72328 	&-4.79341  &2.47 &16.90 &27.34\\
161&36.53788 	&-4.55964  &2.45 &8.64 &11.40\\
162&34.66483 	&-4.69710  &2.43 &50.82 &14.49\\
164&34.62547   &-5.28825   &2.04 &0.19  &2.11\\
165&34.71386 	&-5.15043  &1.75 &16.01 &29.82\\
166&34.50497   &-4.70029   &1.72 &0.109 &1.79\\
\enddata
\tablecomments{$^a$ The spectroscopic redshifts are from \cite{Orenstein2019,Singh2017,Herzog2016}; SDSS DR17 \citep{accetta2022sdss17}; the Dark Energy Spectroscopic Instrument (DESI) survey \citep{2024AJdesi}\\$^b$ Flux density at 20cm from FIRST catalog.\\$^c$ Flux density at $3.4\mu m$ and $3.6\mu m$ from AllWISE and \textit{Spitzer} SEIP catalog.}
\end{deluxetable}

\begin{figure}
    \centering
    \includegraphics[width=1.0\linewidth]{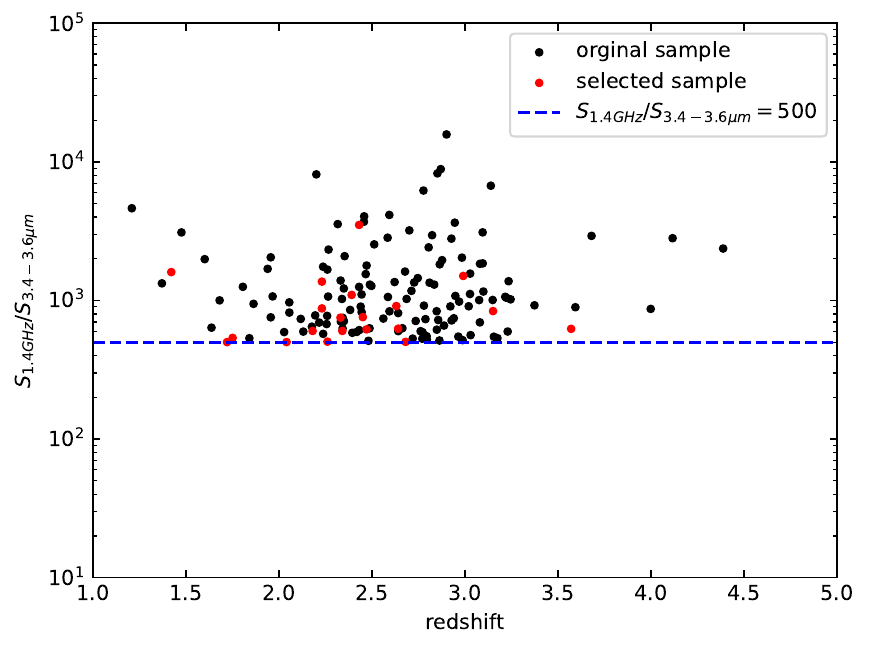}
    \caption{The radio-IR-ratio versus redshift distribution of original sample and selected sample. \textbf{The black dots represent the original IFRS sample. The red dots represent our selected sample of 20 IFRSs for SED modelling. The blue dashed line indicates the selection criteria for IFRS.}}
    \label{fig:enter-label}
\end{figure}

\section{SED modeling} \label{sec:method}
For the SED analysis of our IFRS sample, we used the latest version of the Bayesian SED modeling and interpretation code BayeSED V3.0\citep{han2014bayesed,han2018bayesed,Han2023}\footnote{\url{https://github.com/hanyk/BayeSED3}}. This updated version has been tested on mock galaxy samples and has demonstrated accuracy and speed in estimating galaxy parameters. BayeSED employs principal component analysis (PCA) to reduce the dimensionality of the model SED library. Then, it utilizes an artificial neural network (ANN) or K-Nearest Neighbors (KNN) searching to generate and evaluate the model SED at any point in the parameter space. The MultiNest sampling algorithm is used to obtain the posterior probability distribution of the parameters and Bayesian evidence of the SED model used.

To understand the physical mechanism responsible for their extreme radio-to-IR flux density ratios, we need to determine the relative contributions of different components. The stellar emission is modeled by using the \cite{bruzual2003stellar} simple stellar population (SSP) library with the \cite{chabrier2003imf} initial mass function (IMF), an exponentially declining star formation history (SFH), and the \cite{calzetti2000dust} dust attenuation law. The free parameters in stellar component modeling include stellar age, SFH declining timescale $\tau$, stellar metallicity $Z$ and dust attenuation $A_{V}$. The IR emission could come from hotter AGN-heated dust and/or colder star-formation-heated dust. We assumed an energy balance between stellar emission and cold dust emission, where the stellar emission absorbed by cold dust was completely re-emitted in the IR regime. The cold dust emission was modeled by a graybody (GB), which was defined as
\begin{equation}
    S_{\lambda}\propto(1-e^{-(\frac{\lambda_{0}}{\lambda})^{\beta}})B_{\lambda}(\lambda, T_{dust})
    \label{eq:gb}
\end{equation}

where $\lambda_0=125\ \mu m$ and $B_{\lambda}$ is the Planck blackbody spectrum. The free parameters of the GB model are the dust temperature $T_{dust}$ and the emissivity index $\beta$. The AGN component is independently modeled using the CLUMPY torus model \citep{nenkova2008torus1,nenkova2008torus2}\footnote{{\url{https://www.clumpy.org/}}}, which is given as an extensively sampled library of AGN SEDs. The CLUMPY torus model has six free parameters: the ratio of the outer to inner radii $Y=R_0/R_d$, the observer's line-of-sight inclination $i$, the radial density profile index $q$ ($\rho \propto r^{-q}$), the clump's optical depth $\tau_V$, the number of clouds along a radial equatorial path $N_0$ and the angular distribution width parameter $\sigma$. The CLUMPY model includes not only the torus dust emission but also part of the AGN accretion disk emission, thus providing a consistent modeling of the UV-to-millimeter SED of AGN. We assumed that the priors of the 12 free parameters followed truncated uniform distributions. The prior ranges are summarized in Table \ref{tab:priors}. 

\begin{deluxetable*}{ccccc|cc|cccccc}
\tablecaption{Summary of the free parameters and priors}
\tabletypesize{\scriptsize}
\tablecolumns{13}
\tablewidth{0pt}
\label{tab:priors}
\tablehead
{\\
  & \multicolumn{4}{c}{SSP} &\multicolumn{2}{c}{GB}& \multicolumn{6}{c}{CLUMPY}}
\startdata
Parameters & $\log(age/yr)$ & $\log(\tau/yr)$ & $\log(Z/Z_{\odot})$ & $A_v/mag$ & $T_{dust}/K$ & $\beta$ & $N_0$ & $Y$ & $i$ & $q$ & $\sigma$ & $\tau_\nu$ \\
Prior range & [5, 10.3] & [6, 12] & [-2.3, 0.7]& [0, 4]& [10, 100]& [1, 3]& [1, 15]& [5, 100]& [0, 90]& [0,3]& [15, 70]& [10, 300]\\
\enddata
\end{deluxetable*}

\section{Results}\label{sec:results}
\subsection{Model Comparison}
\begin{figure}[h]
    \centering
    \subfigure[]{
    \includegraphics[width=.5\textwidth]{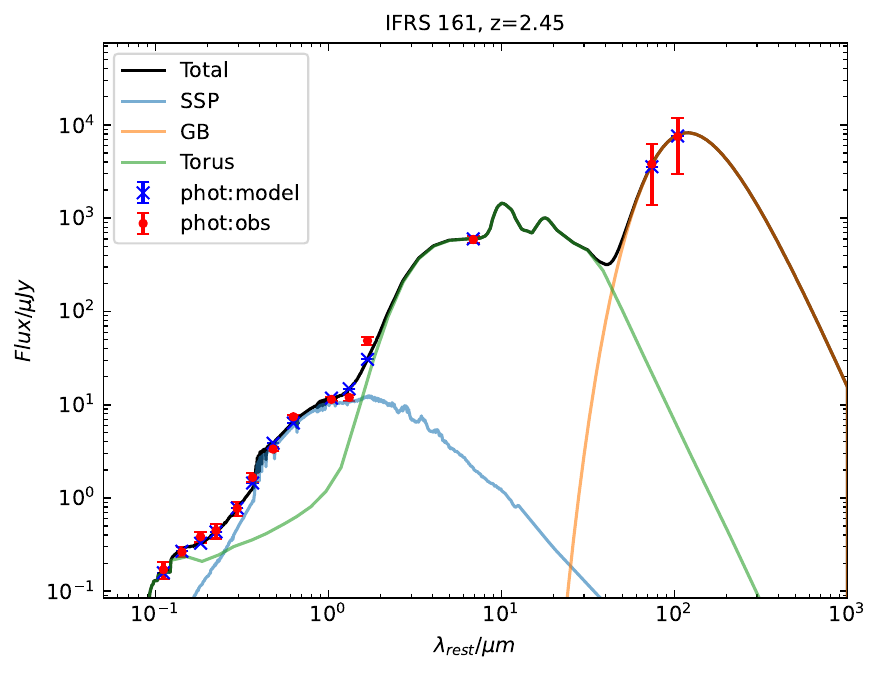}}\subfigure[]{
    \includegraphics[width=.5\textwidth]{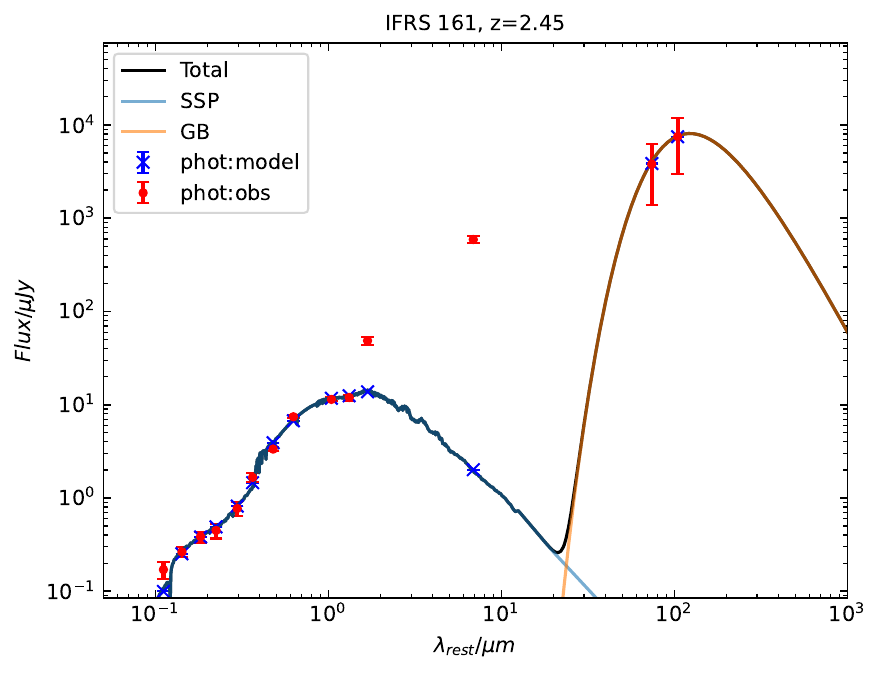}
    }
    \caption{Left: the three-component (SSP+Torus+GB) model fit of IFRS 161. Right: the two-component (SSP+GB) model fit of IFRS 161. The red points with error bars represent the observed SED and the black solid lines represent the total model fit. The blue, green, and orange dashed lines represent the emissions from stars, AGN and dust, respectively.}
    \label{fig:model_compare}
\end{figure}

Of the 20 sources on which we conducted the SED fitting, 6 sources were detected in \textit{Herschel}/SPIRE bands, where cold dust emission dominates. To determine the presence of an AGN component in IFRS, we performed two-component (SSP+GB) and three-component (SSP+Torus+GB) SED modeling on the six IFRSs. Fig \ref{fig:model_compare} shows the best-fitting results of IFRS 161 with and without the AGN torus model. In the case of IFRS 161, the SSP+Torus+GB model seems to provide a better fit to the observations than the SSP+GB model. However, the SSP+Torus+GB model also introduces more free parameters than the SSP+GB model. According to the principle of Occam's razor, a simpler model with a compact parameter space should be preferred over a more complicated one, unless the latter can provide a significantly better explanation of the data\citep{han2018bayesed}. To compare different models quantitatively, we chose to use Bayesian evidence rather than the Bayesian Information Criterion (BIC) because Bayesian evidence provides a more comprehensive framework. Bayesian evidence integrates over the entire parameter space and accounts for uncertainty in parameter estimates and prior distributions, whereas BIC only offers asymptotic approximations to Bayesian evidence. Furthermore, Bayesian evidence is better suited for comparing complex models, such as our multicomponent SED models, especially with a relatively small sample size. The SSP+Torus+GB model will have lower Bayesian evidence unless it provides a significantly better fit than the SSP+GB model. 

\begin{figure}[!htbp]
    \centering
    \includegraphics[width=1.0\textwidth]{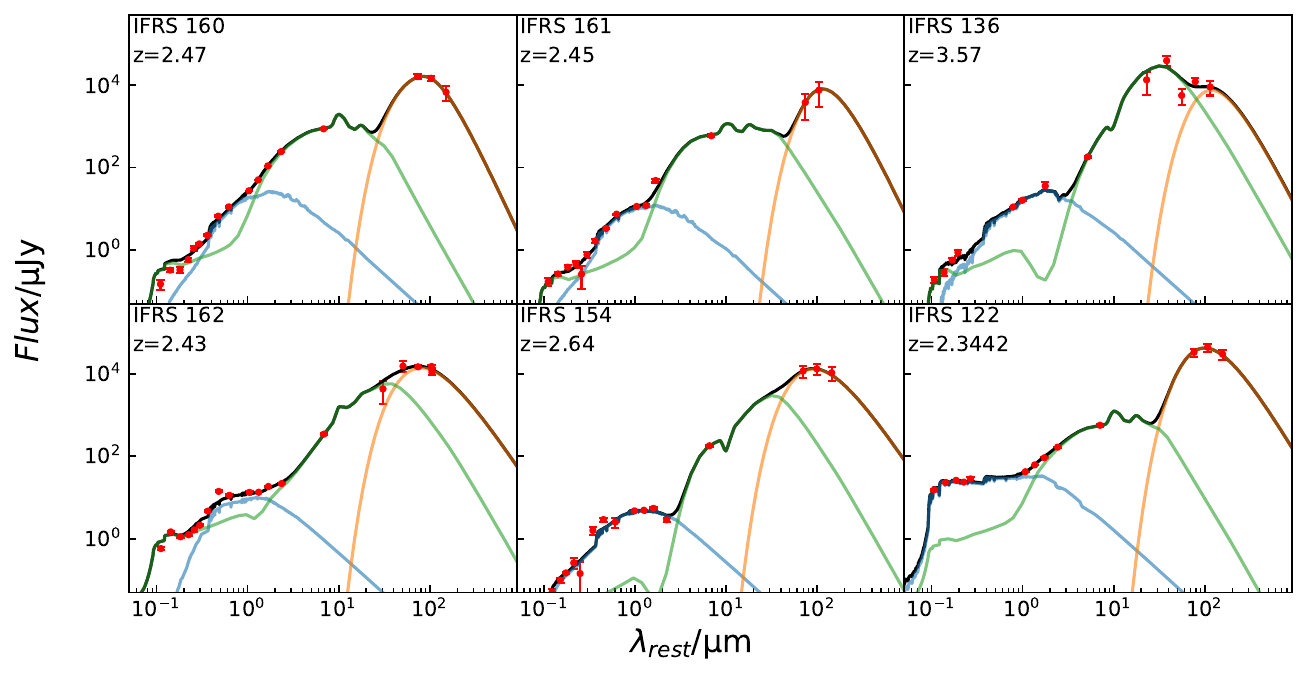}
    \caption{Best-fit three-component model SEDs for 6 IFRS in our sample. The source ID and redshift are shown in each panel. The red points with error bars represent the observed photometric data. The blue, green, and orange solid lines represent the emissions from stars, AGN and dust components, respectively. The solid black line represents the total SED.}
    \label{fig:6bestfits}
\end{figure}

In Table \ref{tab:baye_evi}, we present the natural logarithm $ln(ev_{AGN})$, $ln(ev_{no\ AGN})$ of Bayesian evidence and the Bayes factor $ln(\frac{ev_{AGN}}{ev_{no\ AGN}})$ for the two models. We find that the SSP+Torus+GB model has significantly higher Bayesian evidence than the SSP+Torus model for all IFRSs in the subsample. According to the empirically calibrated Jeffrey scale \citep{jeffreys1998theory,trotta2008bayes}, $ln(\frac{ev_{AGN}}{ev_{no\ AGN}})>5$ (corresponding to odds of about 150:1) indicates strong evidence in favor of an AGN torus component in IFRSs. Therefore, we utilize the results of the three-component model in the following discussion. In Figure \ref{fig:6bestfits}, we show the best-fit SEDs of six IFRSs with the three-component models. Absorbed stellar emission, AGN emission, and cold dust emission are represented by the blue, green, and orange dashed lines, respectively. We adopted the median and 68\% confidence intervals of the posterior probability distribution of each parameter as the fiducial value and uncertainties. The comparison of models and the derived properties will be discussed in the next section.

\begin{deluxetable}{lccc}
\centering
\tablecaption{The Bayesian Evidence of SSP+Torus+GB Model and SSP+GB Model}
\label{tab:baye_evi}
\tabletypesize{\scriptsize}
\setlength{\tabcolsep}{2pt}
\tablehead{\\
\colhead{Source ID} & 
\colhead{$ln(ev_{AGN})$} & 
\colhead{$ln(ev_{no\ AGN})$} & 
\colhead{$ln(\frac{ev_{AGN}}{ev_{no\ AGN}})$}\\}
\startdata
\textbf{122} & -76.841$\pm$0.46 & -950.661$\pm$0.422 & 873.77$\pm$0.882\\
\textbf{136} & -88.215$\pm$0.476 & -209.998$\pm$0.351 & 121.783$\pm$0.827\\
\textbf{154} & -68.342$\pm$0.534 & -172.174$\pm$0.339 & 103.832$\pm$0.873\\
\textbf{160} & -108.569$\pm$0.575 & -6939.67$\pm$0.526 & 6831.101$\pm$1.101\\
\textbf{161} & -63.536$\pm$0.386 & -125.898$\pm$0.326 & 62.362$\pm$0.712\\
\textbf{162} & -184.425$\pm$0.486 & -742.109$\pm$0.379 & 557.684$\pm$0.865\\
\enddata

\end{deluxetable}

\subsection{Median SED}
\begin{figure}[h]
    \centering
    \includegraphics[width=.9\textwidth]{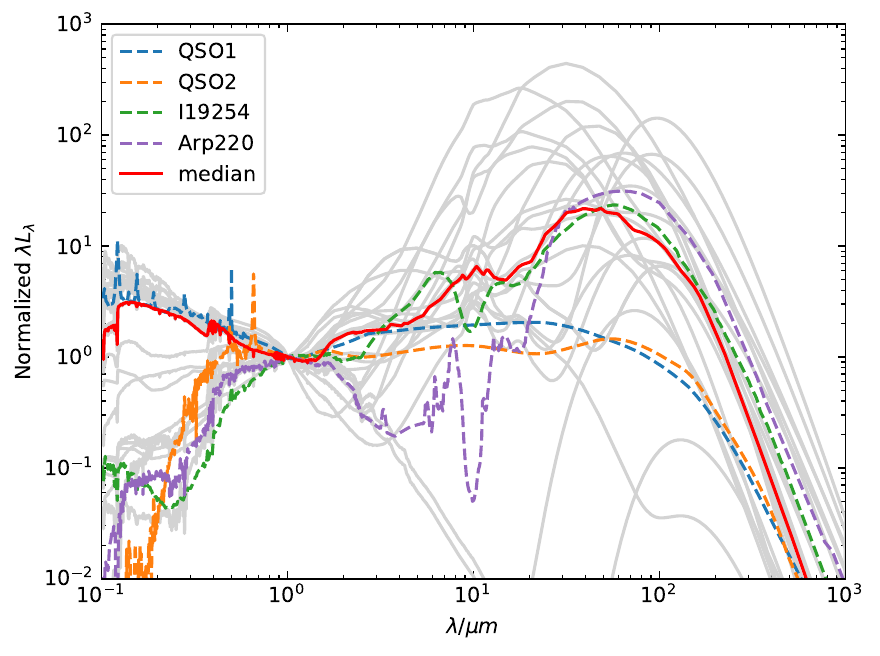}
    \caption{Normalized rest-frame SEDs (gray lines) and the median SED (red solid lines) of 25 IFRS. The SEDs are based on the best-fitting SSP+Torus+GB model and normalized to the luminosity at $\lambda=1\ \mu m$. Individual SEDs and the median SED are compared to templates from \cite{polletta2007spectral}, including Type-1 QSO, Type-2 QSO, starburst galaxy Arp 220 and AGN-starburst composite galaxy IRAS 19254-7245 South (denote as I19254 in the figure). All SED templates are normalized to the luminosity at $\lambda=1\ \mu m$.}
    \label{fig:median}
\end{figure}

In Figure \ref{fig:median}, we present the best-fitting rest-frame SEDs of 20 IFRS (gray lines) using the SSP+Torus+GB model. The SEDs are normalized to the luminosity at $\lambda=1\ \mu m$. Then we derived the median SED of the 20 normalized SEDs of IFRS, represented by the solid red line in Figure \ref{fig:median}. We compare the median SED of the IFRS subsample with some known templates from \cite{polletta2007spectral}, including Type-1 QSO, Type-2 QSO, starburst galaxy Arp 220 and AGN-starburst composite galaxy IRAS 19254-7245 South (hereafter I19254). We find that the median SED of IFRS is similar to that of Type-1 QSO at $0.1\ \mu m <\lambda < 1\ \mu m$. Within the range of $\sim 1-1000\ \mu m$, the median SED is similar to that of I19254. Due to the lack of observational constraints on the FIR SED of most IFRS, the rest-frame SEDs of the IFRS subsample exhibited a large dispersion in the FIR wavelength. 

The normalized rest-frame SEDs of the IFRS sample exhibit two distinctive characteristics within the range of $0.1\ \mu m <\lambda < 1\ \mu m$, showing a flat or a steep spectrum. We then divided the sample into two groups and derived the median SEDs for each group. One median SED shows great consistency with the Type-1 QSO SED throughout the UV-to-IR range, while the other is similar to the I19254 SED. Therefore, we believe that IFRS could consist of at least two distinct populations. Possible reasons for this dischotomy will be discussed in Section \ref{subsec:dichotomy}.

\begin{figure}[!htbp]
    \centering
    \subfigure[]{
    \includegraphics[width=.45\textwidth]{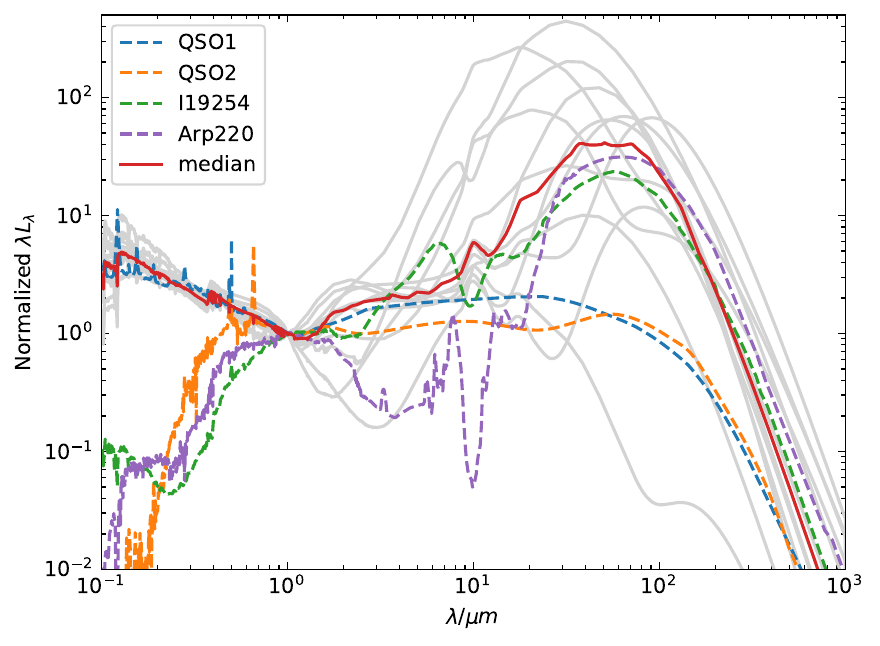}}\subfigure[]{
    \includegraphics[width=.45\textwidth]{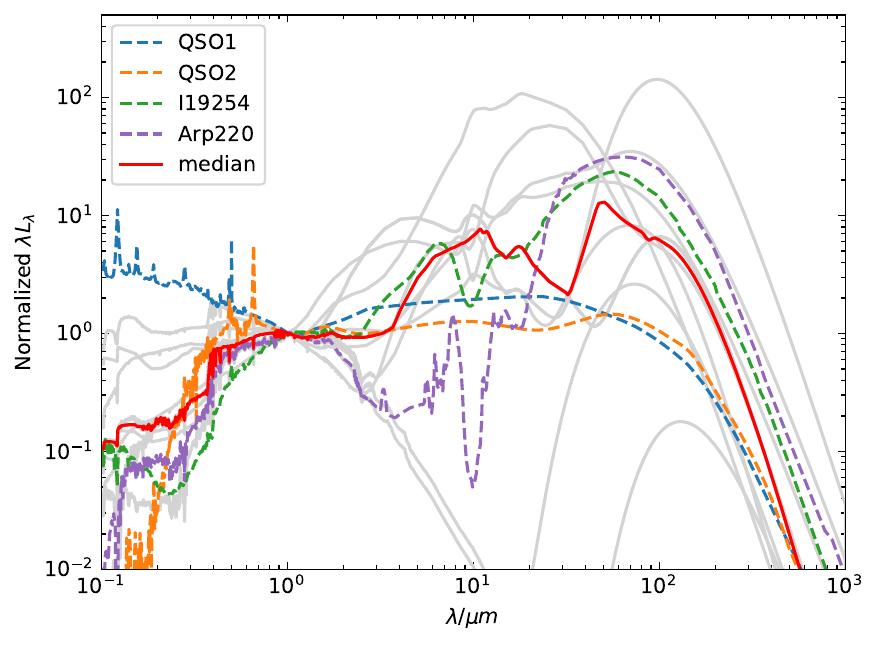}
    }
    \caption{The median SEDs of two distinct groups. Left: median SED of IFRS shows a flat spectrum at $0.1\ \mu m <\lambda < 10\ \mu m$. Right: median SED of IFRS shows a steep spectrum at $0.1\ \mu m <\lambda < 10\ \mu m$.}
    \label{fig:median subsample1}
\end{figure}

\section{Discussion}\label{sec:discussion}

\subsection{SED dichotomy}\label{subsec:dichotomy}
Our study of the IFRS SEDs reveals two distinct groups within our sample. One group exhibits SED characteristics similar to Type-1 QSO, while the other group resembles AGN-starburst composites. Here, we explore the possible physical reasons behind these differences.

The evolutionary stage of the host galaxy and its central AGN can account for the observed SED variations. Type-1 AGN might represent a more evolved state in which the AGN has cleared much of the surrounding dust and gas, revealing the central engine. In contrast, AGN-starburst composites could be in a transitional phase where intense starburst activity is still ongoing and the AGN is actively accreting material, contributing to the composite SED. This scenario suggests that the relatively IR-bright IFRSs might be at an early stage along the AGN-starburst evolutionary sequence.

Differences in the distribution and amount of dust within the host galaxy can lead to distinct SED features. Type-1 AGN typically shows less MIR and FIR emission due to the relatively lower amount of dust along our line of sight. In contrast, AGN-starburst composites have significant dust heated by both AGN and starburst activity, resulting in prominent MIR and FIR emission \citep{2003LonsdaleAGN-SB}. The IR-brighter sources in our sample could result from additional dust obscuration. Variations in dust composition, grain size, and temperature can further modulate the observed SED. The dust properties derived from SED decomposition will be discussed in Section \ref{subsec:dust}.

The intrinsic properties of the host galaxies, such as star formation rate, metallicity, and stellar population age, can also influence the SED. AGN-starburst composites are often found in galaxies with high star-formation rates, leading to strong FIR emission due to heated dust. In contrast, Type-1 AGN hosts may have lower star formation rates, with the AGN dominating the SED\citep{hickox2014sfrBH}. Differences in the stellar population of the host galaxy and the properties of the interstellar medium can therefore contribute to the observed SED dichotomy. Our IFRS sample has a relatively low SFR according to the SED decomposition, which will be further discussed in Section \ref{subsec:SFR}.

While star-forming activities and dust obscuration in the host galaxy may contribute to the SED dichotomy observed in our IFRS sample, we cannot exclude the possibility that differences in the viewing angles of AGN could also contribute to this dichotomy, as one subgroup also bears some resemblance to Type-2 QSO. In Type-2 QSO, the accretion disk and  broad-line region are obscured by the torus, resulting in significant reduction or complete absence of optical and UV emission in the observed SED\citep{1993ARA&Aunimodel}. However, it is important to note that the IFRS subgroup still shows higher IR luminosities compared to Type 2 QSO template, indicating that there are multiple factors contributing to the SED dichotomy, such as the presence of dust. Therefore, the SED dichotomy in IFRS is likely due to multiple factors, including intrinsic properties of host galaxies and viewing angles. 

\subsection{IR-radio relation}
The IR-radio relation is an important indicator of the emission mechanism in galaxies. In SFG, the IR emission mainly originates from the re-emission of dust particles, whereas the non-thermal radio emission primarily arises from synchrotron radiation in the supernova remnants. The IR and radio emissions in SFG show a strong correlation because supernovae typically occur in young stellar populations. Deviations from this correlation, caused by an excess of radio emission, usually imply the presence of RL AGN.  

\begin{figure}[!htbp]
    \centering
    \includegraphics[width=.9\textwidth]{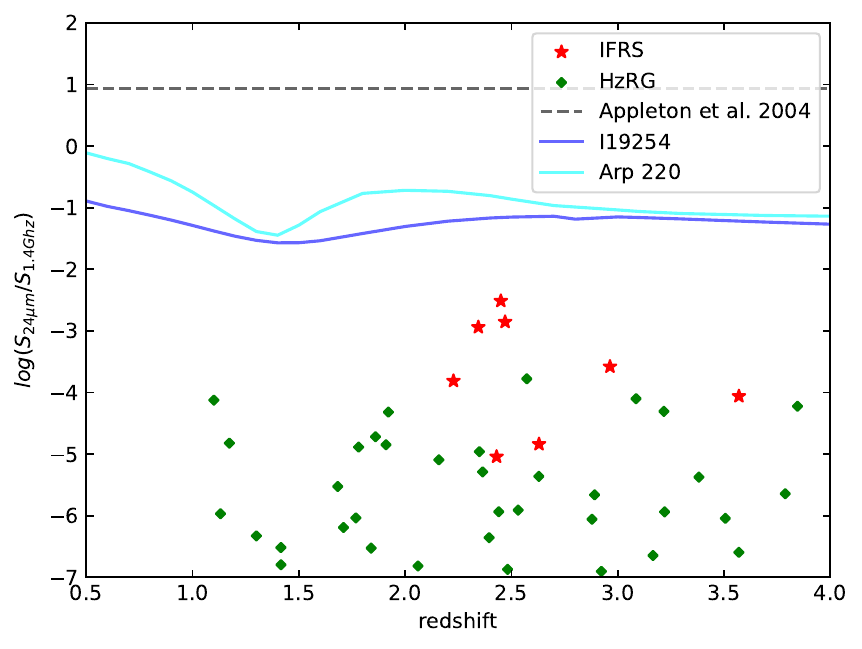}
    \caption{The flux density ratio between $24\mu m$ and $1.4 GHz$ as a function of redshift. The black filled circles represent our IFRS sample and the green crosses represent HzRGs from \cite{seymour2007massive}. The blue solid line and yellow solid line indicate the expected loci of I19254 and Arp 220 \citep{polletta2007spectral}. The red dashed line represents the typical value of $q_{24}\simeq 1.0$ for SFG from \citep{appleton2004far}.}
    \label{fig:IR-radio}
\end{figure}

We investigated the ratio of $24\ \mu m$ MIR to $1.4$ GHz radio flux densities of 8 IFRS with Spitzer/MIPS $24\ \mu m$ observations, which is defined as $q_{24}=\log(S_{\rm 24\mu m}/S_{\rm 1.4GHz})$ \citep{appleton2004far}. The parameter $q_{24}$ is conventionally used to distinguish between the populations of AGN and SFG. In Figure \ref{fig:IR-radio}, we compare the $q_{24}$ of our IFRS as a function of redshifts to that of different types of populations, including starburst galaxies, AGN-starburst composites and HzRGs from \cite{seymour2007massive}. The loci of Arp 220 and I19254 in the diagram $q_{24}-z$ are determined using templates from \cite{polletta2007spectral}. The 8 IFRSs clearly deviate from the IR-radio flux density correlation of the SFGs and partially overlap with the population of HzRGs in Figure \ref{fig:IR-radio}. The HzRGs from \cite{seymour2007massive} have IR luminosity similar to that of Luminous Infrared Galaxies (LIRGs) and Ultra-Luminous Infrared Galaxies (ULIRGs). Therefore, it seems likely that a fraction of the IFRS in this paper represents the extension of the HzRGs at lower luminosities. 

As illustrated in Figure \ref{fig:IR-radio}, some IFRS have $q_{24}$ between SFG and HzRG. Considering the similarity between the SEDs of IFRS and the AGN-starburst composite I19254, it seems likely that a portion of IFRS consists of starburst galaxies with more prominent AGN than I19254. 

\subsection{Star formation rate}\label{subsec:SFR}
We disentangled the IR emissions from AGN-heated warm dust and star formation-heated cold dust in SED modeling.  In Table \ref{tab:derive_properties}, we list the total IR luminosities ($L^{\rm tot}_{\rm IR}$), the IR luminosity of AGN ($L^{\rm AGN}_{\rm IR}$) and the formation of stars ($L^{SF}_{IR}$) within the 8-1000 $\mu m$ range. We found that the six FIR-detected IFRSs in our sample have fairly high IR luminosities ($L^{\rm tot}_{\rm IR}\ >\ 10^{12}L_{\odot}$), all of which exceed the limit of ULIRGs. 

\begin{deluxetable}{lcccc}
\tablecaption{Summary of the derived properties of SED modelling}
\label{tab:derive_properties}
\tabletypesize{\scriptsize}
\setlength{\tabcolsep}{2pt}
\tablehead{\\
\colhead{Source ID} & 
\colhead{$\log(L^{\rm tot}_{\rm IR})$} & 
\colhead{$\log(L^{\rm SF}_{\rm IR})$} & 
\colhead{$\log(L^{\rm AGN}_{\rm IR})$} &
\colhead{$T_{\rm dust}$}\\
 &
\colhead{($L_{\odot}$)}&
\colhead{($L_{\odot}$)}&
\colhead{($L_{\odot}$)}&
\colhead{(K)}\\}
\startdata
\textbf{122}&$12.77^{+0.02}_{-0.03}$&$12.71^{+0.02}_{-0.02}$&$11.89^{+0.06}_{-0.07}$&$39.5^{+1.55}_{-1.52}$\\
\textbf{136}&$13.39^{+0.03}_{-0.01}$&$12.13^{+0.07}_{-0.04}$&$13.37^{+0.02}_{-0.01}$&$26.06^{+5.53}_{-5.18}$\\
\textbf{154}&$12.6^{+0.05}_{-0.03}$&$12.32^{+0.02}_{-0.01}$&$12.28^{+0.07}_{-0.07}$&$43.44^{+0.87}_{-0.73}$\\
\textbf{160}&$12.56^{+0.02}_{-0.01}$&$12.39^{+0.02}_{-0.02}$&$12.07^{+0.01}_{-0.01}$&$50.59^{+1.51}_{-2.13}$\\
\textbf{161}&$12.33^{+0.09}_{-0.07}$&$11.89^{+0.04}_{-0.02}$&$12.13^{+0.12}_{-0.1}$&$29.57^{+2.12}_{-1.48}$\\
\textbf{162}&$12.86^{+0.02}_{-0.02}$&$12.18^{+0.04}_{-0.02}$&$12.76^{+0.01}_{-0.02}$&$33.18^{+2.25}_{-2.07}$\\
\enddata
\end{deluxetable}

\begin{figure}[h]
    \centering
    \includegraphics[width=.9\textwidth]{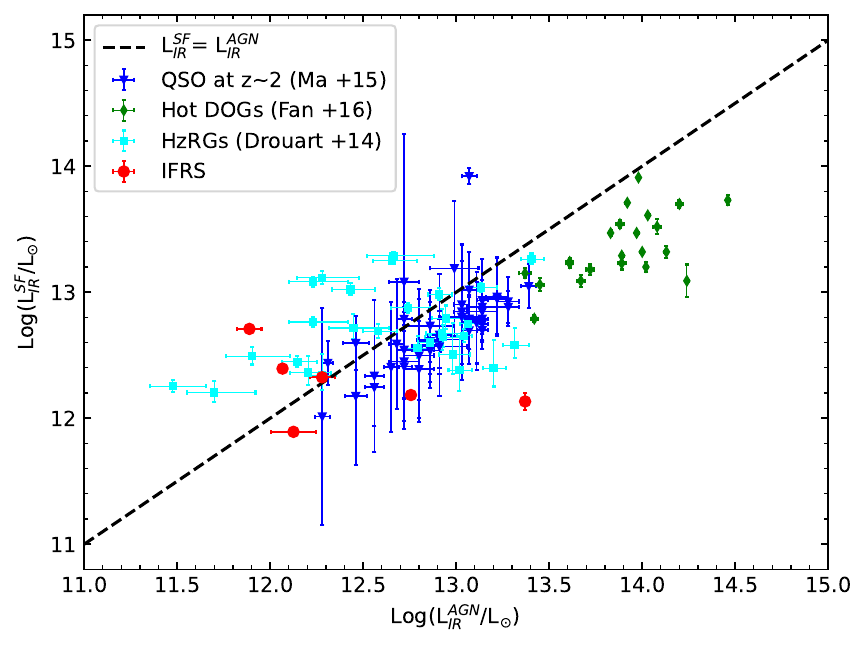}
    \caption{IR luminosity due to star forming activities $L^{\rm SF}_{\rm IR}$ as a function of AGN IR luminosity for our IFRS sample and other populations: Hot DOGs at $z>2$ \citep{fan2016infrared, sun2024}, QSOs at $z \sim 2$ \citep{ma2015co}, HzRGs at $1<z<4$ \citep{drouart2014hzrg}.}
    \label{fig:sf-agn}
\end{figure}

In Figure \ref{fig:sf-agn}, we plot $L_{\rm IR}^{\rm SF}$ versus $L_{\rm bol}^{\rm AGN}$ and compare the 6 FIR detected IFRSs in our sample with different populations, including hot dust-obscured galaxies (Hot DOGs) in $z>2$ \citep{fan2016infrared,sun2024}, QSOs in $z \sim 2$ \citep{ma2015co}, and HzRGs at $1<z<4$ \citep{drouart2014hzrg}. Previous research suggested that IFRSs could be a weaker subgroup of HzRGs \citep{Collier2014,herzog2015dust}. Most of our IFRS sample lies on the low-luminosity end of HzRGs as illustrated by Figure \ref{fig:sf-agn}. 

Given the high IR luminosities of the FIR-detected IFRSs, we will use the simple relation between the star formation rate (SFR) and IR luminosity for local galaxies \citep{kennicutt1998star},
\begin{equation}\label{eq:sfr}
    SFR\ =\ 1.72\times 10^{-10}\times L_{\rm IR}^{\rm SF}
\end{equation}
where $L_{\rm IR}^{\rm SF}$ is in units of $L_{\odot}$ and SFR in $M_{\odot}\ yr^{-1}$. Our FIR-detected sample spans a small range of SFR, from 100 to 900 $M_{\odot}\ yr^{-1}$. Considering the non-detection in the FIR regime for a significant fraction of IFRSs, the SFR estimation of our sample could be an upper limit for the IFRS population. HzRGs span a much larger SFR range, from 100 to $\sim$ 5000 $M_{\odot}\ yr^{-1}$ \citep{drouart2014hzrg}, similar to that of submillimeter galaxies (SMGs) over the same redshift range \citep{wardlow2011laboca,swinbank2014alma}.

The high radio luminosity and extreme radio-to-IR flux density ratio indicate strong AGN activities in IFRSs. However, despite the evidence for significant impact of AGN on host galaxies, the SFR of IFRSs in our sample showed no correlation with the AGN luminosity. It should be noted that the lack of correlation does not rule out a possible relationship between the two parameters because global star formation and AGN activity occur over different timescales, suggesting that estimations of the instantaneous AGN luminosity may not be closely related to its long-term average. Such variations may mask any underlying relationship \citep{hickox2014sfrBH}. Another reason for the observed lack of correlation could be the limited size of our IFRS sample, which can lead to stochastic variations that mask the potential correlation. In the future, we expect a larger IFRS sample with IR detection to testify to this correlation. Furthermore, as the SED dichotomy in Section \ref{subsec:dichotomy} points out, IFRS may represent a heterogeneous population at different evolutionary stages. Some IFRS might be experiencing a surge in AGN activity with declining star formation. This diversity can obscure a clear correlation between AGN luminosity and SFR.

\subsection{Dust properties}\label{subsec:dust}

\begin{figure}[!htbp]
    \centering
    \includegraphics[width=.9\textwidth]{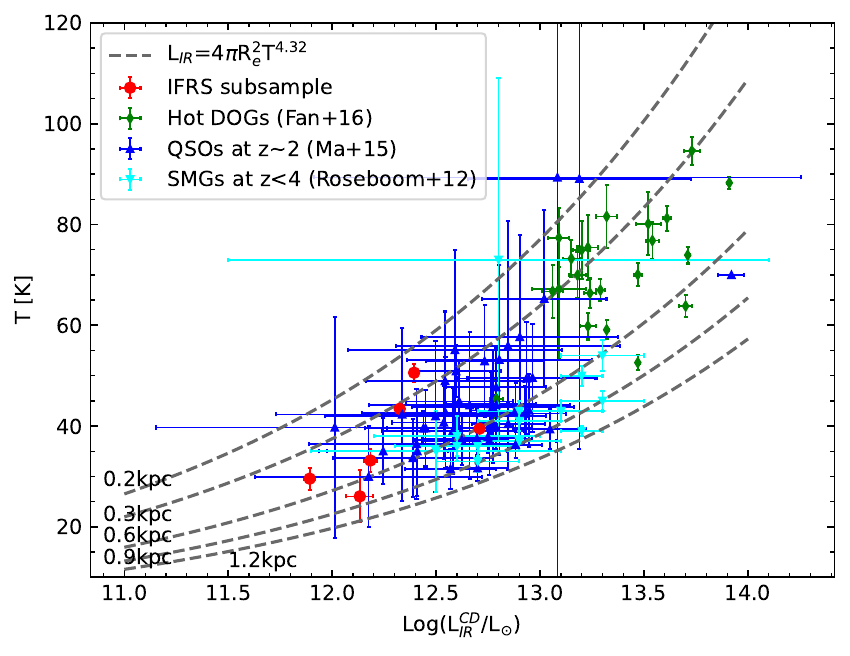}
    \caption{Cold dust temperature as a function of IR luminosity for our IFRS sample and other populations: Hot DOGs at $z>2$ \citep{fan2016infrared, sun2024}, QSOs at $z \sim 2$ \citep{ma2015co}, and SMGs at $z<4$ \citep{roseboom2012herschel}. The grey dashed lines represent $T_{\rm dust}-L_{\rm IR}$ relation with different $R_e$ values (0.2, 0.3, 0.6, 0.9, and 1.2 kpc).}
    \label{fig:LIR-T}
\end{figure}

In Table \ref{tab:derive_properties}, we list the temperature of the cold dust in the IFRS subsample. The dust temperature of the IFRS ranges from 26 to 50 K, with a median value of 36.34 K. The derived dust temperatures are hotter than those found in ULIRGs, SMGs, and DOGs on average, which range from 20 to 50 K \citep{Magdis2012herschel,Melbourne2012spectral}. In Figure \ref{fig:LIR-T}, we plot the relation between the temperature of the cold dust $T_{dust}$ and the IR luminosity of the cold dust. We compare our IFRS sample with other populations: Hot DOGs at $z>2$ \citep{fan2016infrared, sun2024}, QSOs at $z \sim 2$ \citep{ma2015co}, and submillimeter galaxies (SMGs) at $z<4$ \citep{roseboom2012herschel}. The compared samples also used the graybody model described in Equation \ref{eq:gb}. 

The locus of our IFRS sample in the $T_{dust}-L_{IR}$ plane is consistent with that of the QSO sample at $z \sim 2$, but the QSO sample from \cite{ma2015co} is slightly more luminous and has a hotter dust temperature. Recall that for a perfect blackbody, the Stefan-Boltzmann law gives $L=4\pi R^2 \sigma T^4$. According to \cite{ma2015co}, the equivalent for a graybody with general opacity should follow $L_{\rm IR}=4\pi R_e^2\sigma T^{\alpha}$,where $R_e$ could be interpreted as the effective radius of the FIR-emitting region of a galaxy. The value of the index $\alpha$ depends on the choice of the dust temperature range. We adopt the value 4.32 for index $\alpha$ given by \cite{ma2015co} in Figure \ref{fig:LIR-T}. We plot the $T_{\rm dust}-L_{\rm IR}$ relation following the modified graybody equation with a series of different $R_e$ (0.2, 0.3, 0.6, 0.9 and 1.2 kpc; see the gray dashed lines in Figure \ref{fig:LIR-T}). Our IFRS subsample spans a relatively small range in $R_e$ (from 0.3 to 0.6 kpc), indicating that the increase in the IR luminosity of cold dust is mainly due to the increase in the temperature of the dust. 

It is important to note that the sample selection of our analysis targets towards the brighter end of IFRSs. As a result, the derived properties and statistical distributions might be biased, limiting the applicability of these results to the fainter members of the IFRS population. The diversity in the spectral indices among the IFRSs also suggests that they should not be treated as a homogeneous source population. The fainter IFRSs that are undetectable in IR wavelengths may exhibit different physical properties compared to those included in our sample. Future studies incorporating a more comprehensive sample and a deeper IR survey are needed to verify the robustness of the conclusions drawn here.
    
\section{Summary}\label{sec:summary}

In this paper, we present an optical to IR SED analysis of 20 IFRS with spectroscopic redshift at $1<z<4$. We construct the SED by combining SDSS, \textit{Spitzer} IRAC and MIPS, \textit{Herschel} PACS and SPIRE data, and other available optical and IR observations. We used an updated version of BayeSED to model the observed SED and infer the physical properties of our IFRS sample, such as the IR luminosity and dust temperature. Our main results are summarized as follows.

\begin{enumerate}
    \item We compare Bayesian evidence for the three-component model (SSP + Torus + GB) and the two-component model (SSP + GB). We find that the three-component model has more Bayesian evidence than the two-component model for all FIR-detected IFRSs in our sample, indicating that the IFRSs are most likely to be AGN.
    \item We construct a median IFRS SED by taking the median value of 20 normlized rest-frame SED of 20 IFRSs. We categorize our sample into two groups based on their luminosity ratios between optical and mid-IR. The median SEDs of the two groups show similarities with Type-1 QSO and AGN-starburst composite SED templates, respectively, suggesting that IFRSs are made up of at least two different populations. 
    \item  The FIR-detected IFRSs in our sample have a relatively high IR luminosity, with a typical value of $\log L_{\rm IR}\sim12.69^{+0.04}_{-0.03} L_\odot$. We disentangle the IR luminosity of AGN and star formation and find the contributions from the two components to be approximately equal. We estimate the upper limit of the SFR of IFRS to be $\sim 900 M_{\odot}\ yr^{-1}$. The SFRs of our IFRS sample show no correlation to the AGN luminosity, despite a clear indication of AGN impact on host galaxies. The observed dichotomy in SED characteristics is likely due to different evolution stages, with some sources being dominated by AGN activity, while others exhibit significant starburst activity alongside AGN features.
    \item Our IFRS subsample has relatively low dust temperatures, with a typical value of $T_{\rm dust}\sim 36.34^{+1.90}_{-1.79} K$. Adopting the $T_{\rm dust}-L_{\rm IR}$ relation of $L_{\rm IR}=4\pi R_e^2\sigma T^{4.32}$, we find that the IFRS subsample spans a relatively small range in $R_e$, suggesting that the increase in the IR luminosity of the cold dust is mainly due to the increase in the temperature of the dust. 
\end{enumerate}

Future deeper IR surveys will be crucial in providing more comprehensive data on IFRS, enabling more robust conclusions about their physical properties and evolutionary pathways. These surveys will help us better understand the connection between AGN activity and star formation, shedding light on the role of IFRS in galaxy evolution at high redshifts. Enhanced infrared sensitivity and coverage will allow us to detect fainter counterparts and improve the representativeness of our samples, ultimately advancing our knowledge of these enigmatic sources.

\begin{acknowledgments}

We thank the anonymous referees for constructive comments. This work is supported by National Key Research and Development Program of China (2023YFA1608100). LF gratefully acknowledges the support of the National Natural Science Foundation of China (NSFC, Grant Nos. 12173037, 12233008), the CAS Project for Young Scientists in Basic Research (No. YSBR-092), the Fundamental Research Funds for the Central Universities (WK3440000006) and Cyrus Chun Ying Tang Foundations.
YH also gratefully acknowledges support from National Key Research and Development Program of China (Nos. 2021YFA1600401 and 2021YFA1600400), the ``Light of West China'' Program of Chinese Academy of Sciences, the Yunnan Ten Thousand Talents Plan Young \& Elite Talents Project, the Natural Science Foundation of Yunnan Province (No. 202201BC070003), and the International Centre of Supernovae, Yunnan Key Laboratory (No. 202302AN360001).

\end{acknowledgments}

\bibliography{IFRS}{}
\bibliographystyle{aasjournal}

\end{CJK*}
\end{document}